\documentclass[a4paper, 12pt]{article}

\begin{document}

\noindent{\huge\bf Cooper Pairs in Alternating Layers of
Light and Heavy Atoms}
\vspace{4mm}\\
\noindent X. H. Zheng
\vspace{4mm}\\
\em Department of Pure and Applied Physics, The Queen's University of Belfast,
Belfast BT7 1NN, Northern Ireland \em
\vspace{4mm}\\
\noindent E-mail: xhz@qub.ac.uk\\
Tel: 44 (0)28 90 335054\\
Fax: 44 (0)28 90 438918\\
\vspace{7mm}\\
\noindent{\bf\Large Abstract}
\vspace{4mm}\\ 
The Hamiltonian and trial function in the BCS theory are 
improved to test the limit of this theory.
The Cooper pairs arise from standing electron waves,
ready to move with atoms, giving high $T_c$.
The Hamiltonian is derived from 
alternating layers of light and heavy atoms, 
giving a forbidden zone hosting no standing wave pairs.
The exchange term may force singlet pairs into this zone,
leaving triplet pairs outside, giving magnetic excitations.
If the Fermi energy is crossed only by the CuO$_2$ band,
then the forbidden zone and triplet pairs will vanish,
consistent with experimental evidence.
\vspace{70mm}

\newpage
\noindent
Recently there has been considerable interest in the original 
theory of Bardeen, Cooper and Schriefer (BCS)~\cite{BCS}.  
It was used by Nunner, Schmalian and Bennemann 
to explain the isotopic effect of cuprates~\cite{Nunner}.  
It was also used by Nozi\`{e}res and Pistolesi to study pseudogaps~\cite{Nozieres}.  
Following this line of thinking, we adopt the view that
cuprates have a Fermi surface and electron-phonon interactions.  
We also adopt the BCS formalism, which is relatively simple, 
due to neglecting damping and retardation, 
with negligible effects~\cite{Schrieffer}.

The BCS theory is based on variational evaluation of the Fr\"ohlich Hamiltonian,
with the Cooper pairs as the trial function~\cite{BCS,Frohlich}.  
With an improved Hamiltonian and 
trial function, we may explain further properties of cuprates.   
In order to improve the Hamiltonian, we use a simple model of alternating layers of 
light and heavy atoms (Fig.\ 1), because in cuprates the CuO$_2$ layers are always 
sandwiched by heavy atoms.
For example, in La$_2$CuO$_4$ we have a ratio 96/310 when comparing the atomic wt.\ 
of the CuO$_2$ layer and the two LaO layers.  In YBa$_2$Cu$_3$O$_7$ we have a ratio 
281/386 when comparing the two CuO$_2$ layers (including the Y atom in between) 
with other atoms.  In Tl$_2$Ba$_2$CuO$_6$ the ratio is 96/746, which becomes 
256/746 and 416/746 when the number of CuO$_2$ layers are 2 and 3 (including Ca in 
between).  In HgBa$_2$CuO$_4$ the ratios are 96/507, 232/507 and 368/507 
when the number of CuO$_2$ planes are 1, 2 and 3
(including Ca in between)~\cite{Poole}.

The classic Cooper pairs arise from traveling electron waves,
which are mobile in all directions.
However, the resistance of cuprates in the $c$ axis is 
significantly larger than that in the $a$-$b$ plane~\cite{Poole}:
most electrons are mobile only in that plane.
It is reasonable to assume that the atomic layers act as 
potential wells to retain electrons.
Being reflected back and forth in potential wells, 
electrons must be in the form of standing waves.  
Indeed, according to energy band calculation, 
in cuprates the Fermi surfaces join together in the $c$ direction~\cite{Pickett}, 
a familiar sign of electron standing waves in that direction.   
In the case of Bragg scattering only some electrons are reflected,
giving narrow `necks' to join Fermi surfaces.
In cuprates virtually all electrons are reflected in the $c$ direction, 
so that the Fermi surface becomes cylindric.  In fact, many cuprate theories
assume that carriers are somehow bound around the CuO$_2$ plane~\cite{Chakravarty}, 
i.e. they are standing waves across that plane.  
We add two Bloch functions together to model electrons in cuprates.  
The result is a traveling wave in the $a$-$b$ plane but a standing wave
in the $c$ direction (Fig. 1).  The Cooper pairs arise from such waves.  
These pairs concentrate in the layers of light atoms
(CuO$_2$ planes) as a natural result of our theory.  

First, we explore the prediction power of the BCS theory.  The reduced Hamiltonian~\cite{BCS}
involves a series of pair generation and destruction operators with the \em c\em -number coefficient
\begin{equation}
V({\bf k , q} )=\sum_{l=1}^{3}\frac{2\hbar\omega_{l}({\bf q})\mathcal{M}_{l}^2({\bf k,q})}
{[\hbar\omega_{l}({\bf q})]^{2}-[\epsilon({\bf k + q})-\epsilon({\bf k})]^{2}}
\label{eq:c-number}
\end{equation}
\noindent where {\bf k} and $\epsilon$ are electron vector and energy (measured from the 
Fermi surface), {\bf q} and $\omega_{l}$ phonon vector and frequency, and $l$ identifies 
phonon branch (excluding transverse phonons, which do not interact with electrons in 
$N$-processes)~\cite{Ziman}.  The matrix element
\begin{equation}
\mathcal{M}_l({\bf k, q})= \tilde{q}_l\left[\frac{\hbar N}
{2M\omega_l({\bf q})}\right]^{1/2}\int_{\Omega}\psi_{{\bf k+q},\sigma}^{*}({\bf r})
\delta \mathcal{V}({\bf r})\psi_{{\bf k},\sigma}({\bf r})d{\bf r}
\label{eq:matrix}
\end{equation}
\noindent measures the strength of electron-phonon interaction.  
Here $\psi$ is the electron wave function, 
$\sigma$ spin $\uparrow$ or $\downarrow$, \em M \em the mass of an atom, 
\em N \em the number of atoms in unit volume, 
{\bf r} the coordinates in real space, $\Omega_{0}$ a volume surrounding the atom, 
$\Gamma_{0}$ its boundary, 
and $\delta\mathcal{V}({\bf r}) = \mathcal{V}({\bf r}) - \mathcal{V}(\Gamma_{0})$, 
$\mathcal{V}$ being the potential field.  
We define $\tilde{q}_l$ as the \em l\em-th component of $U{\bf q}$, 
$U$ being the $3 \times 3$ unitary matrix found 
when solving the classical equation of motion for the atom.  
Mott and Jones found matrix elements when $\Omega_{0}$ is the Wigner-Seitz cell~\cite{Mott}.  
We find equation~\ref{eq:matrix} when $\mathcal{V}(\Gamma_{0})$ is constant 
(this defines $\Omega_{0}$ in a natural manner). The BCS self-consistent equation~\cite{BCS}
\begin{equation}
\Delta({\bf k})=\sum_{\bf q}V({\bf k, q})\frac{\Delta({\bf k + q})}
{[\Delta^{2}({\bf k + q}) + \epsilon^{2}({\bf k + q})]^{1/2}}
\label{eq:self-consistent}
\end{equation}
\noindent is an integral equation of the Cauchy type:
$V({\bf k, q})$ is singular~\cite{Porter}.

We solve equation~\ref{eq:self-consistent} through iteration~\cite{Porter}.  With a proper first 
approximation, we may expect reasonable accuracy after just one iteration.  We use free electron energy to 
evaluate equation~\ref{eq:c-number}, where the denominator turns out to be $4\epsilon_{F}\epsilon_{\bf q}
[\delta_{l}^{2} - (\zeta + \cos\theta)^{2}]$, $\epsilon_{F} = (\hbar^{2}/2m)|{\bf k}|^{2}$ is the Fermi
energy (we study $|{\bf k}|$ near the Fermi surface), $\epsilon_{\bf q} = (\hbar^{2}/2m)|{\bf q}|^{2}$,
$\delta_{l}^{2} = (m/2)v_{l}^{2}/\epsilon_{F}$, $v_{l} = \omega_{l}({\bf q})/{\bf q}|$ the sound velocity.  
In the Debye approximation $\delta_{l} = (Z/16)^{1/3}\Theta_{D}/\Theta_{F} \approx 10^{-3}$ in all 
superconducting metals, where $Z$ is the valency, $\Theta_{D}$ and $\Theta_{F}$ are the Debye and Fermi 
temperatures.  Therefore $V({\bf k, q}) > 0$ (condition to have an energy gap) holds in equation~\ref{eq:c-number} 
only when $\zeta + \cos\theta \approx 0$, $\zeta = |{\bf q}|/|2{\bf k}|$, $\theta$ being the angle
between {\bf k} and {\bf q}, so that $|{\bf k + q}|^{2} = |{\bf k}|^{2} + |{\bf q}|^{2} + 2|{\bf k}|
|{\bf q}|\cos\theta \approx |{\bf k}|^{2}$.  Thus $\epsilon({\bf k + q}) \approx \epsilon({\bf k})$, i.e.\
electrons changes direction but not energy in scattering, which is used as our first approximation
(also used to study metal resistivity)~\cite{Mott}.  The use of free electron energy implies a spherical
Fermi surface and hence an isotropic energy gap, so that $\Delta({\bf k + q}) = \Delta(|{\bf k + q}|)
\approx \Delta(|{\bf k}|) = \Delta({\bf k})$.  These lead through equation~\ref{eq:self-consistent} to:
\begin{equation}
E_{TRV}=\sum_{\bf q}V({\bf k,q}) = \frac{\hbar e^{2}}{k_{B}T_{\rho}}\eta\rho nv^{2}
\label{eq:0Kgap}
\end{equation}
\noindent where $E_{TRV} = [\Delta^{2}({\bf k}) + \epsilon^{2}({\bf k})]^{1/2}$ at $T = 0$, $TRV$ standing
for travelling wave, $e$ and $n$ are electron charge and density, $k_{B}$ the Boltzmann constant,
$\rho$ the resistivity at temperature $T_{\rho}$, $v = k_{B}\Theta_{D}/\hbar k_{D}$ the Debye sound
velocity, $k_{D}$ being the phonon cut-off wavenumber, and
\begin{equation}
\eta = \frac{1}{\pi}\int_{0}^{(4Z)^{-1/3}}F^{2}(x)\frac{\zeta^{2}d\zeta}{1 - \zeta^{2}}
\bigg/\int_{0}^{(4Z)^{-1/3}}F^{2}(x)\zeta^{3}d \zeta \approx 1
\label{eq:eta}
\end{equation}
\noindent Here $F(x) = 3(x \cos x - \sin x)/x^{3}$ is the overlap integral function, $x = 3.84\alpha^{1/3}
Z^{1/3}\zeta$, $\alpha = N \Omega _{0}/\Omega$ the fraction of $\Omega _{0}$ in a primitive cell, and 
$\Omega$ the unit volume.  We assume ${|\bf q}|/|2{\bf k}| = \zeta <(4Z)^{-1/3} < 1$, because 
equation~\ref{eq:c-number} arises from a canonical transformation~\cite{Frohlich}, where operator
commutation requires ${\bf q}\neq \pm 2{\bf k}$.  In first iteration, the pair occupancy varies linearly 
if $-E_{TRV} < \epsilon({\bf k}) < E_{TRV}$, otherwise equation~\ref{eq:self-consistent} has improper solutions 
(occupancy $< 0$ or $> 1$).  This justifies the BCS approach to integrate equation~\ref{eq:self-consistent}
only in a thin layer across the Fermi surface~\cite{BCS}.  This surface does not have to be 
spherical, so long as $k_D << |{\bf k}|$ (integration area small).

In order to find $T_c$ we follow BCS~\cite{BCS} to minimize the free energy
of the electron-phonon system. 
Evaluating the result via iteration, we find
\begin{equation}
E = E_{TRV}\tanh(E/2k_{B}T)
\label{eq:gapfunction}
\end{equation}
\noindent where $E = [\Delta^{2}({\bf k}) + \epsilon^{2}({\bf k})]^{1/2}$ at $T > 0$.  Since 
$\tanh(E/2k_{B}T) < E/2k_{B}T$, we have $T \leq E_{TRV}/2k_{B}$ and hence $2E_{TRV}/k_{B}T_{c} =4$.
It is easy to prove, by direct substitution, that $E$ from equation~\ref{eq:gapfunction} is also
the \em exact \em solution of equation 3.27 in~\cite{BCS} (BCS self-consistent equation for $T > 0$),
provided that $\hbar\omega N(0)V = E_{TRV}$ in that equation.  Clearly, $E$ is not a function of 
{\bf k}, the so-called gap parameter $\Delta({\bf k})$ is, contrary to general perception.  
This is justified: according to BCS it is $E$ that measures the energy gap, $\Delta$ is just an
approximation, i.e.\ $\epsilon\approx 0$ near the Fermi surface giving $E = (\Delta^2 + 
\epsilon^2)^{1/2}\approx\Delta$~\cite{BCS}.  What arises from experiment is actually 
$2E_{TRV}/k_{B}T_{c}$, whose value may deviate from 4 for reasons other than $\hbar\omega/k_{B}T_{c}
>> 1$ (weak coupling).  Since $E$ is constant, any Cooper pairs are equally likely to be 
excited.  This is also justified: $\mathcal{M}_{l}$ (measuring the strength of 
electron-phonon interaction) in equation~\ref{eq:matrix} varies little across the Fermi surface.

According to equations~\ref{eq:0Kgap} and~\ref{eq:gapfunction} 
a good superconductor must have numerous free electrons
(large $n$) scattered frequently by atoms (large $\rho$) moving quickly to facilitate pairing 
(large $v$).  The factor $\eta$ arises when the summation over {\bf q} in equation~\ref{eq:gapfunction}
is replaced by an integration over $(4\pi/3)k_{D}^{3}$ (volume of the first Brillouin zone), which 
exists in the sense of the Cauchy principal value (used by Kuper to verify the BCS 
theory)~\cite{Porter, Kuper}, i.e.\ positive and negative contributions of $V({\bf k, q})$, if finite,
are cancelled on a series of spherical surface, the singular point ignored.  We are entitled to do
so, because equation~\ref{eq:c-number} is defined on a grid of {\bf k} and {\bf q}, which may not be in
precise combinations to let $V({\bf k, q}) = \infty$.  We can also avoid such combinations by
suppressing a few phonons with little physical consequence.  This principal value varies little among
phonon branches, allowing us to use $\sum_{l}\tilde{q}_{l}^{2} = {\bf q}^{t}U^{t}U{\bf q} = 
|{\bf q}|^{2}$ ($U$ unitary) to find the numerator in equation~\ref{eq:eta}.  We use the expression for 
metal resistivity to calibrate $\delta\mathcal{V}$, and this leads to the denominator in 
equation~\ref{eq:eta}~\cite{Mott}.  When $\alpha = 1$, equation~\ref{eq:0Kgap} yields 
$2E_{TRV} = 2.2$, 18 and 27 for
Cd, Ta and Nb (1.5, 14 and 30.5 experimentally, in $10^{-4}$eV), which are of the right order,
although over and under-estimtions are possible.  On average equation~\ref{eq:0Kgap} yields $2E_{TRV} 
= 15.7$ for Zn, Cd, Hg, Al, Ga, Tl, Sn, Pb, V, Nb, Ta and Mo (11.3 experimentally).

Now consider a crystal of alternating layers of light and heavy atoms (Fig.\ 1).
For Cooper pairs of traveling electron waves, 
equations~\ref{eq:0Kgap} and~\ref{eq:gapfunction} are still valid.  
The derivation is straightforward in principle but involved technically.  
In equation~\ref{eq:eta} the upper limit of integration is replaced by
$(8Z)^{-1/3}$: the first Brillouin zone is smaller the larger the
primitive cell.  We also have $\alpha = 2N\Omega_0/\Omega$ (assuming
$\Omega_0$ invariant in the cell) and $Z$ averaged over different atoms.
In order to model electrons in cuprates, we notice that, in principle, 
electrons of any configuration can be expanded into Fourier series
in terms of plane waves.  These series can be shortened, when
the plane waves are replaced by waves resembling more closely the actual 
configuration of the electrons in the crystal, as is done by various sophisticated
methods to calculate the electron band structure~\cite{Pickett}.  For simplicity, 
we consider series of superpositions of just two Bloch functions:
\begin{equation}
\psi^{(1)}_{{\bf k},\sigma}({\bf r})\propto\exp[i(k_{x}x + k_{y}y)]\cos(\pi z/c)
\label{eq:standing-wave-1}
\end{equation}
\begin{equation}
\psi^{(2)}_{{\bf k},\sigma}({\bf r})\propto\exp[i(k_{x}x + k_{y}y)]\sin(\pi z/c) 
\label{eq:standing-wave-2}
\end{equation}
\noindent where ${\bf k} = {\bf x}k_x + {\bf y}k_y$ is a 2D wavevector.
Both $\psi^{(1)}$ and $\psi^{(2)}$ are traveling waves in the 
$a$-$b$ plane.  On the other hand, $\psi^{(1)}$ is confined in layer 1 
(its anti-nodes are in that layer), whereas $\psi^{(2)}$ is confined in
layer 2 (Fig.\ 1).  Apparently, electrons overlap with those in the neighboring 
layer, but not with those in the next neighboring 
layer (there is a node in between, see Fig.\ 1).  
These appear to be reasonable as first order approximation.  
We assume that both $\psi^{(1)}$ and $\psi^{(2)}$ are close to the Fermi surface.
This is true at least in some cuprates: 
the chain band crosses $\epsilon_F $ in YBa$_2$Cu$_3$O$_7$, 
so does the TlO band in Tl$_2$Ba$_2$CuO$_6$, the BiO band 
in Bi$_2$Sr$_2$CaCu$_2$O$_8$, etc.~\cite{Pickett}. 

We use standing electron waves to build Cooper pairs.  The formalism 
parallels that for the classic Cooper pairs:
in second quantization the basis states do not have to be plane 
waves~\cite{BCS, Waldram}.  This is necessary, because superconductivity is a 
second order process.  Unless the configuration of (single) electrons is chosen 
properly (a first order process), the energy gap cannot be maximized.
In equation~\ref{eq:c-number}, $V$ is replaced by $V_{ij}$ $(i, j = 1, 2)$, $l$ runs 
from 1 to 6, and $\mathcal M_{l}$ is replaced by $\mathcal M^{(ij)}_{l}$ to link 
$\psi^{(i)}$ and $\psi^{(j)}$.  Both intra $(i = j)$ and inter-layer $(i \neq j)$ 
couplings are possible, because electrons in neighboring layers overlap (Fig.\ 1).  
If we use $c/2$ to replace $c$ in equations~\ref{eq:standing-wave-1} 
and~\ref{eq:standing-wave-2}, then we are out of the first Brillouin zone.  
As a result, $\psi^{(1)}$ and $\psi^{(2)}$ become identical, giving   
no inter-layer coupling and reduced energy gap, as will be shown. 
If we use $2c$, $3c$, ... to replace $c$, then $E_{12}$ in Appendix becomes smaller.
This weakens the inter-layer coupling, a choice not preferred.  Indeed, the classic Cooper 
pair (spin $\uparrow$, $\downarrow$ and wavevector ${\bf k}$, ${\bf -k}$) is also a 
choice (i.e.\ a trial function) to maximize the energy gap.

Letting $f^{(1)}_{\bf k}$ and  $f^{(2)}_{\bf k}$ be the over-all probability of excitations in layer 1
and 2, we have $-4k_{B}\sum_{\bf k}[(f^{(1)}_{\bf k}/2 + f^{(2)}_{\bf k}/2) + (1 - f^{(1)}_{\bf k}/2
- f^{(2)}_{\bf k}/2)\ln(1- f^{(1)}_{\bf k}/2 - f^{(2)}_{\bf k}/2)]$ as the entropy of the pair ensemble,
which is from the consideration that pairs in either layer may have the same energy, so that 
thermodynamically they fall into the same group of entities, i.e.\ there can be 4 electrons at the same 
energy level, giving the degeneracy factor 4.  Note that in cuprates bands 
of both the light and heavy layers may cross $\epsilon_F$~\cite{Pickett}, 
so that the above degeneracy is possible.
Minimizing the free energy of the pair ensemble, we find
through iteration two equations with the solution
\begin{equation}
1-2f^{(1)}_{\bf k} = \frac{E_{12} - E_{22}}{E^{2}_{12} - E_{11}E_{22}}
k_{B}T \ln\frac{2 - f^{(1)}_{\bf k} - f^{(2)}_{\bf k}}{f^{(1)}_{\bf k} + f^{(2)}_{\bf k}}
\label{eq:probability1}
\end{equation}
\begin{equation}
1-2f^{(2)}_{\bf k} = \frac{E_{12} - E_{11}}{E^{2}_{12} - E_{11}E_{22}}
k_{B}T \ln\frac{2 - f^{(1)}_{\bf k} - f^{(2)}_{\bf k}}{f^{(1)}_{\bf k} + f^{(2)}_{\bf k}}
\label{eq:probability2}
\end{equation}
\noindent Here $E_{ij} = \sum_{\bf q}V_{ij}({\bf k, q})$ and the summation over 
${\bf q} = {\bf x}q_{x}+{\bf y}q_{y}$ is in 2D, i.e. the standing electron waves 
emit and absorb phonons in 2D.  Since $E_{ij} = N^{-1}_{z}\sum E_{ij}$ when the 
summation is over $q_{z}$ (not an argument of $E_{ij}$), $N_{z}$ being the number 
of $q_{z}$ in the first Brillouin zone, we integrate $E_{ij}$ in 3D
(in the sense of the Cauchy principal value) for convenience.    
Equations~\ref{eq:probability1} 
and~\ref{eq:probability2} apply to pure intra-layer couplings when $E_{12} = 0$, 
and pure inter-layer couplings when $E_{11} = 0$ and $E_{22} = 0$. 
Both lead to an energy gap, because $E_{ij} > 0$ always holds (Appendix).

The inter and intra-layer couplings are in competition.  
If none dominates, then both are suppressed.  
Since $M_1 < M_2$, we have $E_{22} < E_{11}$ (Appendix).  
When $E_{22} < E_{12} < (E_{11}E_{12})^{1/2}$, 
the two sides of equations~\ref{eq:probability1} have opposite signs: 
$f^{(1)}_{\bf k} = 1/2$ and $T = 0$ must hold to give $f^{(2)}_{\bf k} = 1/2$ 
via equation~\ref{eq:probability2}.  Similarly $T = 0$ when
$(E_{11}E_{22})^{1/2} <E_{12} < E_{11}$.  Clearly $E_{22} < E_{12} < E_{11}$ 
(inter-layer coupling weaker than the coupling in layer 1 but stronger 
than that in layer 2) is a forbidden zone hosting no standing wave pairs.  
Outside this zone, equations~\ref{eq:probability1} and~\ref{eq:probability2} 
can be added together to recover equation~\ref{eq:gapfunction}, with $E$ 
replaced by $1 - f^{(1)}_{\bf k} - f^{(2)}_{\bf k}$ and $E_{TRV}$ replaced by
\begin{equation}
E_{STD} = \frac{E^{2}_{12} - E_{11}E_{22}}{E_{12} - (E_{11} + E_{22})/2}
\label{eq:newgap}
\end{equation}
\noindent $STD$ standing for standing wave.  If all Cooper pairs are excited, 
then $f^{(1)}_{\bf k}, f^{(2)}_{\bf k} \rightarrow 1/2$ giving 
$2E_{STD}/k_{B}T_{c} = 4$ to estimte $T_{c}$. 

In Fig.\ 1 $E_{STD}/E_{TRV} > 5.57$ when $E_{11} < E_{12}$. 
This large ratio arises from equation~\ref{eq:newgap}, where 
$E_{11}\rightarrow E_{12}$ leads to $E_{STD}\rightarrow 2E_{11}$: 
the energy gap is larger the stronger the inter-layer coupling.  
At this point, equation~\ref{eq:probability1} yields $f^{(2)}_{\bf k} = 1/2$: 
all the pairs in layer 2 are excited, 
apparently draining much of the excitation energy.  
Pairs in layer 1 (light atoms) are more or less left alone: superconducting 
carriers are in the CuO$_2$ layers.  Furthermore, $E_{11} > E_{TRV}$ holds 
as a result of the symmetry of the standing waves (with respect to $a$-$b$ 
planes, rather to sites of atoms) reflecting the fact that bound electrons
are readier to move with the atoms.  In Fig. 1 $E_{11} < E_{12}$ 
for $Z < 0.1314$.  This small valency arises, because on average phonons 
have smaller $|{\bf q|}$ to pair electrons on a smaller Fermi sphere, 
so that $\mathcal{M}^{(11)}_{l} (\propto |{\bf q}|) < \mathcal{M}^{(12)}_{l} 
(\propto 2\pi/c)$ holds to give $E_{11} < E_{12}$.  
Assuming $E_{STD}/E_{TRV} = 5.57$, we have 
$T_{c}\approx$ 130K when $E_{TRV} = 40.2\times 10^{-4}$eV 
($40\times 10^{-4}$eV for Nb$_{3}$Ge).  In Ba$_{2}$YCu$_{3}$O$_{7}$ we have 
$k_{D} = (6\pi^{2}N/\Omega)^{1/3} = 6.99\times 10^{9}$m$^{-1}$, $\Theta _{D}\approx 400$K, 
$v = k_{B}\Theta _{D}/\hbar k_{D}\approx 7.49 \times 10^{3}$ms$^{-1}$, 
$\rho = 70-550 \times 10^{-8}\Omega$m ($a$-$b$ plane) and   
$n \approx 6 \times 10^{27}$m$^{-3}$ from infrared reflectivity~\cite{Poole, Thomas}.
Taking surface values, these lead through equation~\ref{eq:0Kgap} to 
$E_{TRV} = 10-77\times 10^{-4}$eV ($\eta = 1$).
A point to notice: $\rho$ is for traveling waves in equation~\ref{eq:0Kgap}.  
Although standing electron waves are easier to be scattered, giving larger $\rho$,
this is more or less compensated by the weak coupling at their nodes.  
It is interesting that, in Ba$_{2}$YCu$_{3}$O$_{7}$, $N = 5.76\times 10^{27}$m$^{-3}$, 
so that $Z = (1/13)n/N \approx 0.08$~\cite{Poole}.
   
We may have spin singlet pairs in the forbidden zone, triplets outside.  
Specifically, while $E_{12} < E_{11}$ holds for singlet pairs, 
$E_{12} > E_{11}$ may hold for triplet pairs, 
i.e.\ $E_{11}$ declines faster when the pair symmetry changes, 
due to the stronger effect of the exchange term on $E_{11}$, 
which is related to intra-layer coupling, 
where electron waves overlap to a greater extent.  
Outside the forbidden zone, $E_{STD}$ changes $\sim$10\% 
when $Z$ drops just 0.001 from 0.1314.  Across the zone border, 
$Z_{STD}$ changes more dramatically (Fig. 1).  
In both cases doping may obscure the isotopic effect.  
The Fermi surface of cuprates is not strictly cylindric~\cite{Pickett}:
classic Cooper pairs may arise to 
give superconductivity inside the forbidden zone.
Since $E_{TRV}$ changes slowly with $Z$ (Fig. 1),
the isotopic effect will be more apparent.  
Indeed, in cuprates the isotopic effect is minimum 
when $T_{c}$ peaks with proper doping~\cite{Franck}. 
Although travelling electron wave pairs (giving $E_{TRV}$) cannot compete with 
standing wave pairs (giving $E_{STD} > E_{TRV}$) in the $a$-$b$ plane, they can 
move in the $c$ axis (standing waves cannot)
to make the Knight shift complicated~\cite{Takigawa, Annett1},
and give energy gap anisotropy and pair symmetry anisotropy
(well documented for cuprates)~\cite{Collins, Tamasaku, Sun, Tsuei, 
Kouznetsov, Wei}.  

In conclusion, rather surprisingly and significantly, the BCS  
theory may play a major role to explain the properties of cuprates.
We show that this theory can be used to calculate $T_c$ 
from first principles, the first time to our knowledge,   
giving $T_c \sim 130$K for cuprates.  
The impression that $T_c$ is low in the BCS theory 
arises from McMillan's calculation, where the $T_c$ 
of an alloy family is clamped to that of the parent metal,
assumed to be a natural element of low $T_c$~\cite{McMillan}.
Another worry about the BCS theory is the Migdal instability
which, according to Waldram, may not set in at 
$130$K~\cite{Waldram, Migdal}.
We also show that in a complex system like cuprates the 
microscopic physics may manifest itself as a paradox:
the exchange term may drive the singlet pairs 
into the forbidden zone, leaving the triplet pairs outside.
Indeed neutron scattering does exhibit a magnetic 
peak below $T_c$ from YBa$_2$Cu$_3$O$_8$ and 
Bi$_2$Sr$_2$CaCu$_2$O$_8$~\cite[and the references therein]{Fong}.
In addition to adding another explanation to the already long
list of explanations, our theory has a specific experimental basis.
In the above cuprates $\epsilon _F$ is crossed
by both the CuO$_2$ and heavy atom layer bands~\cite{Pickett}: 
our theory applicable.  On the other hand, magnetic excitations are 
absent in La$_{2-x}$Sr$_x$CuO$_4$~\cite{Kastner} where $\epsilon _F$ 
is crossed only by the CuO$_2$ band~\cite{Pickett}.
Therefore we have to assume $M_1 = M_2$: heavy atom layers are not 
involved in the electron-phonon interaction near the Fermi surface.
As a result, both the forbidden zone and triple pairs vanish.
It appears worthwhile to search magnetic excitations from
e.g.\ HgBa$_2$Ca$_2$Cu$_3$O$_8$ and Tl$_2$Ba$_2$CuO$_6$,
as example and counter-example of cuprates,
where $\epsilon _F$ is crossed only by the 
CuO$_2$ band~\cite{Pickett, Singh}.
\vspace{6mm}\\

\noindent{\bf\Large Acknowledgement}
\vspace{4mm}\\
\noindent The author thanks Professor C.\ L.\ S.\ Lewis for helpful comments.
\vspace{6mm}\\

\noindent{\bf\Large Appendix}
\vspace{4mm}\\
\noindent For a spherical Fermi surface
\begin{eqnarray}
E_{11} = C \int^{(8Z)^{-1/3}}_{0}\frac{1}{2}\left[\frac{(A + B)^{2}}{M_{1}}
+ \frac{(A - B)^{2}}{M_{2}}\right]\frac{\zeta^{2}d \zeta}{1 - \zeta^{2}}\nonumber\\
\nonumber\\
E_{12} = C \int^{(8Z)^{-1/3}}_{0}\frac{1}{2}\left[\frac{B^{2}}{M_{1}}
+ \frac{B^{2}}{M_{2}}\right]\frac{0.65}{Z^{2/3}}\frac{a^{3}}{c^{3}}
\frac{\zeta^{2}d \zeta}{(1 - \zeta^{2})^{2}}\ \ \nonumber
\end{eqnarray}
\noindent $A$ and $B$ are values of the overlap integral function $F(x)$, with
$x = 3.84 \alpha^{1/3}Z^{1/3}$ $\zeta(1 - \zeta^{2})^{1/2}$ and 
$3.84\alpha^{1/3}[Z^{2/3}\zeta^{2}(1 - \zeta^{2}) + 0.65(a/c)^{4/3}]^{1/2}$, respectively, and 
$C = 6Zm\alpha^{2}(\delta V)^{2}/\epsilon _{F}$.  If $M_{1}$ and $M_{2}$ are interchanged, then 
$E_{11}$ is turned into $E_{22}$.  For Cooper pairs of traveling waves $B = 0$ and 
$x = 3.84\alpha^{1/3}Z^{1/3}\zeta$, so that $E_{11}$ is reduced to $E_{TRV}$ in 
equation~\ref{eq:0Kgap} ($C$ expressed in $n$. $\rho$ and $v$).  
The factor $(1 - \zeta^{2})^{-1}$ in
$E_{11}$ (or $E_{TRV}$) and $E_{12}$ is related to the Cauchy principal value, which starts to fail
when $\zeta = |{\bf q}|/|2{\bf k}|\rightarrow 1$, where the canonical transformation also fails.
However, in Fig.\ 1 $E_{TRV}$ shows little sign of divergence when $Z > 0.13$.  Another factor 
$(1 - \zeta^{2})^{-1}$ in $E_{12}$ is from the denominator ($\propto |{\bf q}|^{2}$) in 
equation~\ref{eq:c-number}, which is cancelled in $E_{11}$ ($\mathcal{M}^{11}_{l}\propto |{\bf q}|$)
but not in $E_{12}$ ($\propto 2 \pi/c$), so that in Fig.\ 1 $E_{STD}$ turns upwards when
$Z \rightarrow 0.13$.
\vspace{6mm}\\

\noindent{\bf\Large Figure legend}
\vspace{4mm}\\
\noindent Fig.\ 1 Crystal of light (wt.\ $M_{1}$ in layer 1, open circles) and heavy atoms
(wt.\ $M_{2}$ in layer 2), $a, b$ and $c$ are lattice constants, the anti-nodes of
$\psi^{(1)}_{{\bf k},\sigma}$ and $\psi^{(2)}_{{\bf k},\sigma}$ are in layer 1 and 2,
respectively; $E_{STD}$ (solid line, a.u.) and $E_{TRV}$ (broken line, a.u.) are found when $M_1/M_2 = 0.7$
and $a = b = 0.5c$; $Z < 0.1314$ ($E_{11} < E_{12}$) and $Z > 0.1361$ ($E_{12} < E_{22}$)
border the forbidden zone.


\newpage

\begin{picture}(400,300)(20,0)

\thinlines
\put(0,0){\framebox(400,300)}

\multiput( 25,  0)(25,0){15}{\line(0,1){5}}
\multiput( 25,295)(25,0){15}{\line(0,1){5}}
\multiput(  0, 30)(0,30){ 9}{\line(1,0){5}}
\multiput(395, 30)(0,30){ 9}{\line(1,0){5}}

\put(-17,-30){\large\sffamily 0.130}
\put( 83,-30){\large\sffamily 0.132}
\put(183,-30){\large\sffamily 0.134}
\put(283,-30){\large\sffamily 0.136}
\put(383,-30){\large\sffamily 0.138}

\put(-30, -5){\large\sffamily 0.0}
\put(-30, 55){\large\sffamily 0.2}
\put(-30,115){\large\sffamily 0.4}
\put(-30,175){\large\sffamily 0.6}
\put(-30,235){\large\sffamily 0.8}
\put(-30,295){\large\sffamily 1.0}

\put(150,-70){\Large VALENCY(Z)}

\thicklines
\qbezier(0,247.89)(32,230.85)(72,217.32)
\put(72,217.32){\line(0,-1){217.32}}
\put(72,0){\line(1,0){232}}
\put(304,0){\line(0,1){130.2}}
\qbezier(304,130.2)(352,125.73)(400,121.2)

\multiput(0,41.61)(8,-0.1866){51}{\circle*{2}}

\put(330, 50){\Large $E_{TRV}$}
\put(330,145){\Large $E_{STD}$}

\thinlines
\multiput(130,260)(40,0){2}{\circle*{5}}
\multiput(115,245)(40,0){2}{\circle*{5}}
\multiput(130,180)(40,0){2}{\circle*{5}}
\multiput(115,165)(40,0){2}{\circle*{5}}

\multiput(115,245)(15,15){2}{\line(1,0){40}}
\multiput(115,245)(40, 0){2}{\line(1,1){15}}
\multiput(115,165)(15,15){2}{\line(1,0){40}}
\multiput(115,165)(40, 0){2}{\line(1,1){15}}

\multiput(130,220)(40,0){2}{\circle{7}}
\multiput(130,140)(40,0){2}{\circle{7}}
\multiput(115,205)(40,0){2}{\circle{7}}
\multiput(115,125)(40,0){2}{\circle{7}}

\multiput(118.5,205)(15,15){2}{\line(1,0){33}}
\multiput(118.5,125)(15,15){2}{\line(1,0){33}}
\multiput(117.475,207.475)(40, 0){2}{\line(1,1){10.05}}
\multiput(117.475,127.475)(40, 0){2}{\line(1,1){10.05}}

\multiput(130,223.5)(40,0){2}{\line(0,1){36.5}}
\multiput(115,208.5)(40,0){2}{\line(0,1){36.5}}

\multiput(130,143.5)(40,0){2}{\line(0,1){73}}
\multiput(115,128.5)(40,0){2}{\line(0,1){73}}

\multiput(90,165)(0,80){2}{\line(1,0){10}}
\multiput(95,165)(0,55){2}{\line(0,1){25}}

\put(108,257){\Large $a$}
\put(146,268){\Large $b$}
\put(92,200.5){\Large $c$}

\put(185,222){\Large 1}
\put(185,182){\Large 2}

\put(215,140){\line(0,1){120}}
\qbezier(215,180)(250,220)(215,260)
\qbezier(215,180)(200,160)(195,140)

\put(270,140){\line(0,1){120}}
\qbezier(270,140)(235,180)(270,220)
\qbezier(270,220)(285,240)(290,260)

\put(192,100){\Large $\psi^{(1)}_{{\bf k},\sigma}$}
\put(255,100){\Large $\psi^{(2)}_{{\bf k},\sigma}$}

\end{picture}

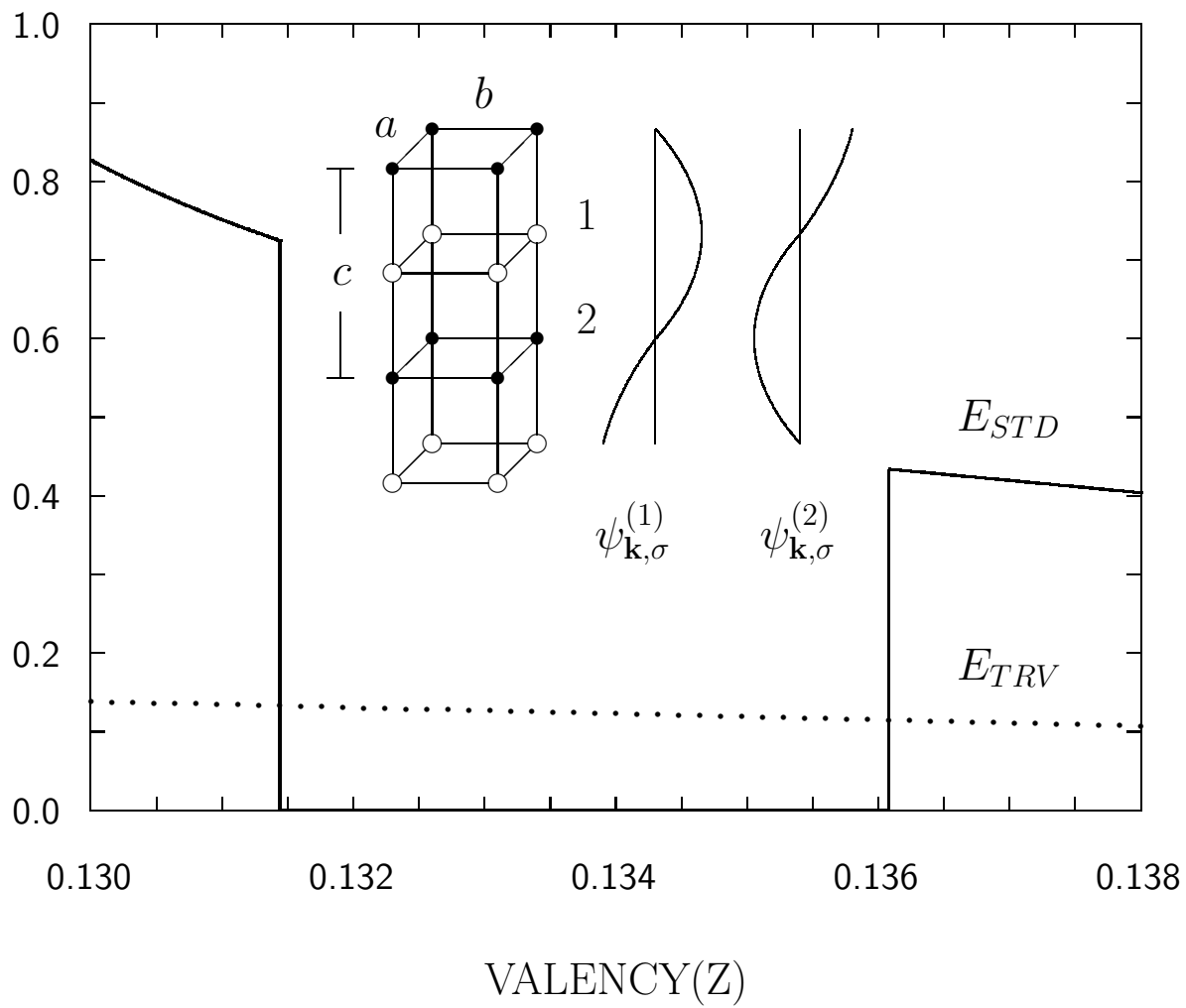
\end{document}